# Nanoscale confinement and control of excitonic complexes in a monolayer WSe$_2$

Hyowon Moon[1,2,*,†], Lukas Mennel[1,3,*], Chitraleema Chakraborty[1,4], Cheng Peng[1], Jawaher Almutlaq[1], Takashi Taniguchi[5], Kenji Watanabe[5], and Dirk Englund[1,†]

[1]Department of Electrical Engineering and Computer Science, Massachusetts Institute of Technology, Cambridge, MA, USA
[2]Center for Optoelectronic Materials and Devices, Korea Institute of Science and Technology, Seoul, Republic of Korea
[3]InstVienna University of Technology, Vienna, Austria
[4]Department of Material Science and Engineering, University of Delaware, DE, USA
[5]National Institute for Materials Science, Tsukuba, Ibaraki, Japan

Corresponding authors: †hwmoon@kist.re.kr; †englund@mit.edu

**Abstract**

Nanoscale control and observation of photophysical processes in semiconductors is critical for basic understanding and applications from optoelectronics to quantum information processing. In particular, there are open questions and opportunities in controlling excitonic complexes in two-dimensional materials such as excitons, trions or biexcitons. However, neither conventional diffraction-limited optical spectroscopy nor lithography-limited electric control provides a proper tool to investigate these quasiparticles at the nanometer-scale at cryogenic temperature. Here, we introduce a cryogenic capacitive confocal optical microscope (C$^3$OM) as a tool to study quasiparticle dynamics at the nanometer scale. Using a conductive atomic force microscope (AFM) tip as a gate electrode, we can modulate the electronic doping at the nanometer scale in WSe$_2$ at 4K. This tool allows us to modulate with nanometer-scale confinement the exciton and trion peaks, as well a distinct photoluminescence line associated with a larger excitonic complex that exhibits distinctive nonlinear optical response. Our results demonstrate nanoscale confinement and spectroscopy of exciton complexes at arbitrary positions, which should prove an important tool for quantitative understanding of complex optoelectronic properties in semiconductors as well as for applications ranging from quantum spin liquids to superresolution measurements to control of quantum emitters.

A central goal in solid-state physics and devices is to control optoelectronic properties and processes at the scale of elementary quanta, which would unlock new applications in optoelectronics, quantum devices, and other fields[1–4]. Layered transition metal dichalcogenides (TMDs) provide a promising platform to achieve this goal as they naturally confine charge carriers in atomically thin two-dimensional (2D) lattices with unique properties[5–7]. Patterned gate electrodes on top of stacked 2D materials allow in-plane electrostatic doping underlying a wide range of optoelectronic devices[8–11] and new possibilities for controlling exciton transport[12,13]. However, reaching the ultimate limit of control at the scale of individual excitonic complexes requires new approaches in electrostatic gating at near the nanometer-scale, since the exciton Bohr radius in TMDs is an order of magnitude smaller than current electric potential resolution[14,15].

An alternate approach is the use of nanoscale probe tip, which achieves extremely localized gating as in scanning tunneling microscopy[16,17]. Here, we report on active low-temperature local doping of 2D semiconductors by using a conductive AFM tip to electrically probe material characteristics in a confocal cryogenic microscope apparatus. This "cryogenic capacitive confocal optical microscope" ($C^3OM$) allows nanometer-scale gating around the tip to be applied anywhere *in-situ*. Our electrostatic modeling indicates that the doping full-width-half-maximum (FWHM) resolution ($r_{dop}$) is determined largely by the spacing ($d$) from the tip to the 2D semiconductor and can even be much smaller than the tip radius $r_{TIP}$, dropping below 1 nm at $d$ = 0.3 nm (a monolayer hBN) and $r_{TIP}$ ~ 10 nm. In experiments performed on $WSe_2$ under $d$ = 10 nm of hBN, the tool allows deterministic modulation of the exciton and trion peaks, as well as a distinct spectroscopic line associated with a larger excitonic complex, which we observe a nonlinear relationship between optical pump power and photoluminescence. Combined with the ability for local strain application[18,19], the C³OM provides a tool for nanometer-scale control of quasiparticle dynamics at the nanometer scale.

As illustrated in Fig. 1a, the C³OM consists of a transparent substrate positioned over a scannable AFM cantilever inside a closed-cycle liquid helium cryostat. A scanning confocal microscope images through the substrate to the sample, which consists of an hBN-encapsulated WSe$_2$ monolayer and a multilayer graphene back-gate. Two metal electrodes connected to the WSe$_2$ and graphite control the electric potential of the sample ($V_S$) and the back gate ($V_{BG}$), respectively. The Cr- and Pt-coated conductive AFM tip has a radius of 25 nm and biases local WSe$_2$ via a conductive cantilever of a voltage $V_{TIP}$. We assembled the sample by dry transfer from exfoliated flakes (see Materials and Methods); Fig. 1b shows the fabricated sample.

Figure 2a and 2b map the photoluminescence (PL) intensity of the monolayer WSe$_2$ flake for $V_S$ = 0 V and $V_{BG}$ = -3 V under continuous-wave (CW) optical excitation (wavelength: 633 nm, power: 3 µW), when the tip in contact with the top hBN (white circles). As the tip voltage ramps from $V_{TIP}$ = 0 V (Fig. 2a) to $V_{TIP}$ = 10 V (Fig. 2b), the PL around the tip is strongly modulated. Figure 2c plots the spectra when $V_{TIP}$ increases from 0 V (black) to 10 V (brown) ($V_S$ = $V_{BG}$ = 0 V). At $V_{TIP}$ = 0 V, the spectrum shows the typical WSe$_2$ PL emission with the exciton ($X^0$), negatively charged trion ($X^-$), and a broad localized exciton band ($X^{LB}$). At $V_{TIP}$ = 10V, a peak at 1.675 eV dominates the spectrum; we associate this with the $X^{-'}$ spectroscopic feature[20–24] previously reported under strong electron doping in W-based TMDs and attributed to different excitonic complexes such as the fine structure of negative trions[20], doubly negatively charged excitons ($X^{2-}$) [21], and many-body interaction between excitons and short-wave intervalley plasmons[25–27]. In previous experiments with a large (>µm) static back-gate, this feature appears in the PL when $n_e$ > 2 x 10$^{12}$ cm$^{-2}$ and dominates it when $n_e$ > 5 x 10$^{12}$ cm$^{-2}$ while the exciton and trion peaks disappear[27]. Figure 2d shows that the bright spot follows the tip position when we move it laterally, although there is a sample inhomogeneity due to the transfer process. Therefore, we can perform our experiment in a relatively clean and homogeneous region after the final fabrication process of the sample.

Figure 3a-c shows the gate voltage-dependent PL emission, which displays the spectra when $V_{TIP}$ sweeps from -10 to 10 V in the 1 V step as the $V_{BG}$ changes from 1 to -1 V in the -1V step. Figure 3d-f plots the Lorentzian-fitted PL intensity of each peak. When $V_{BG} = 1$ V (Fig. 3a,d), $X^-$ peak at 1.705 eV dominates the signal when $V_{TIP} < 0$ V, and a new peak at lower energy appears at $V_{TIP} = 2$ V. The intensity of the peak becomes stronger while its energy decreases as $V_{TIP}$ increases until both saturate at $V_{TIP} > 6$ V. The reduced emission from the $X^-$ peak suggests that the formation of the new peak is related to $X^-$. We do not find a significant change in the localized band. When $V_{BG} = 0$ V (Fig. 3b,e), we observe both $X^-$ and $X^0$, because the sample is negatively doped, naturally. Supplementary Fig. 1 confirms that the charge neutral point is achieved at $V_{BG} = -0.3$ V by sweeping $V_{BG}$ from -3 to 3V in the 0.1 V step while keeping $V_{TIP}$ at the ground. $X^-$ peak maximizes at $V_{TIP} = 4$ V, and then reduces again as $X^{-'}$ dominates at higher $V_{TIP}$. We observe $X^0$ and positively charged trion ($X^+$) peaks at $V_{BG} = -1$ V (Fig. 3c,f), while $X^{-'}$ appears again at high $V_{TIP}$. By comparing Fig. 3a-c, we observe: (i) the onset voltage of $X^{-'}$ increases as $V_{BG}$ decreases; (ii) $X^{-'}$ emission intensity increases as $V_{TIP}$ or $V_{BG}$ increases; (iii) $X^0$, $X^-$, $X^+$ emissions do not fully disappear only by $V_{TIP}$, but they do by $V_{BG}$. (i) and (ii) suggests that the carrier concentrations in the material play a crucial role in the $X^{-'}$ emission, while (iii) arises from the confined nature, which we discuss in the next paragraph with the theoretical calculation.

We calculated the capacitance, total charges, electric fields, and corresponding distribution of charges and electrons in the monolayer $WSe_2$ with and without photogenerated electron-hole pairs at various excitation powers (see Methods, Supplementary Fig. 2). In the calculation, the AFM tip is approximated as a conductive sphere of radius $a$ = 25 nm, which is apart from the surface of WSe$_2$ flake by $d_Z$ (Supplementary Fig. 2). The maximum electron density ($n_{e,max}$) in the WSe$_2$ flake and the half-width-half-maximum (HWHM) confinement radius $r_C$ strongly depend on the distance $d_Z$ (Supplementary Fig. 3) and the radius $a$ (Supplementary Fig. 4). Figure 4a shows the lateral distribution of particle density when a continuous-wave laser (wavelength: 633 nm, power: 3 µW) generates electron-hole pairs in the WSe$_2$. $V_{TIP}$ increases from 0 (gray) to 10 V (vivid) in the 2 V step, and $d_Z$ is set to 10 nm, which is the AFM-measured thickness of the top hBN. The naturally doped electron density is 2 x 10$^{11}$ cm$^{-2}$, calculated from the thickness of the bottom hBN (17 nm) and $V_{BG}$-sweep experiment (Supplementary Fig. 1). The dotted yellow line shows the density of photogenerated electron-hole pairs (n$_P$), calculated from the laser intensity distribution approximated as a Gaussian function[28]. The exciton (n$_X$) and trion (n$_T$) densities are calculated from the mass action model[29,30], which is

$$\frac{n_e n_X}{n_T} = \frac{4 M_X m_e}{\pi \hbar^2 M_T} k_B T \, exp(-\frac{E_T}{k_B T}), \tag{1}$$

where $k_B$ is the Boltzmann constant, $T$ is the temperature, $E_T$ is the trion binding energy, $M_X$ = $m_e + m_h$ and $M_T = 2m_e + m_h$ are the exciton and trion effective masses, and $m_e$ and $m_h$ are the electron and hole masses. These simulations indicate that higher $V_{TIP}$ increases $n_e$ and $n_T$, while suppressing n$_X$ at the center. The confinement radii are calculated as 22 nm and 153 nm for electrons and trions, smaller than the excitation beam radius of 195 nm. The excitons are confined in a ring with inner and outer radii of 68 nm and 267 nm, respectively. The normalized particle distribution in the 2D map is shown in Fig. 4b (scale bar: 200 nm).

Our calculation shows that $r_C$ can be an order of magnitude smaller than $r_{TIP}$ (Supplementary Fig. 4b). In particular, by reducing $d_Z$ to the thickness of a monolayer of hBN (0.33nm), a relatively large tip of $r_{TIP}$ = 15 nm can confine the electrons to a disk of radius of 3 nm, which is comparable to the Bohr radius of negative trions or higher excitonic states [31,32] (Supplementary Fig. 5). $V_{TIP}$ also modulates the effective radius of the heavily doped region of $n_e > 5 \times 10^{12}$ cm$^{-2}$, where the X$^-$ peak dominates the PL signal[27], at the nanometer scale .

Finally, we study the photophysical characteristics of the confined X$^-$ emission by photoluminescence (PL), at λ=633 nm and P=3 μW focused to the tip. In the following experiments, $V_{TIP}$ and $V_{BG}$ are kept at 10 V and 0 V, respectively. The X$^-$ feature shows circular polarization but not linear polarization (Supplementary Fig. 6a,b), which shows that this peak is more trion-like than exciton-like [20]. A photoluminescence excitation resonance at the neutral exciton energy of 1.74 eV confirms that the peak originates from the free-exciton population in WSe$_2$ (Supplementary Fig. 6c).

Figure 4c plots the pump-power dependence of the confined X$^-$ emission intensity, $I_P$. We fit this measurement by the model $I_P = C_P \cdot P^\alpha$, where the $C_P$ is the proportional constant, $P$ is the pump power, $α$ is the exponent. The value of $α$ varies for different excitonic complexes; for example $α$ is close to 1 for excitons ($α_X \sim 1$) and trions ($α_T \sim 1.15$), while it is close to 2 for neutral biexcitons ($α_{XX} \sim 1.94$) and negatively charged biexcitons ($α_{XX^-} \sim 1.82$)[33]. Figure 4d shows that below 9 μW, X$^-$ peak increases linearly with $α_{X^-} = 0.99 \pm 0.01$. This suggests that the X$^-$ emission originates from additional charges with a single exciton rather than from biexcitonic states or higher-order excitonic complexes. However, as the power exceeds 9 μW, we observe an extraordinary behavior: X$^-$ peak saturates and decreases fast ($α = -1.27 \pm 0.24$). In contrast, trion and exciton emissions grow sublinearly ($α_T = 0.90 \pm 0.01$ and $α_X = 0.50 \pm 0.06$) at low power, but jump with $α_T = 2.21 \pm 0.25$ and $α_X = 3.15 \pm 0.03$ at high power. We attribute the superlinear increase of excitons and trions to the dissociation of X$^-$ into X$^-$ and X$^0$ due to the excessively supplied electron-hole pairs supplied by the strong excitation power. However, this behavior has not been observed in the delocalized X$^-$ peak at a similar or higher excitation power[20,24], and the calculated maximum density of photogenerated holes is two orders of magnitude smaller than that of electrons. This suggests that the strong electron confinement promotes the dissociation of X$^-$ peak at a much lower excitation power due to the diffusion of photogenerated holes from the surrounding region, as the photogeneration occurs in an order of magnitude larger area than the confined electrons. Supplementary Fig. 7 plots the ratio of the total number of photogenerated holes ($N_P$) to that of electrons ($N_e$) within the integration radius $r_i$, confirms that the total number of holes within the radius of 200 nm is over 6% of that of electrons, which is an order of magnitude larger than the ratio at the center, 0.45%.

In conclusion, we presented a method to control and monitor excitonic complexes in a monolayer semiconductor at nanoscale. Highly confined electric fields induced by a conductive AFM tip generate strongly confined excitonic complexes near the tip position, with the possibility of confinement to nanometer-scale doping profiles of disks or rings for thinner hBN layers. The electrostatic gating experiment allows control of position, on/off, and the emission energy of the excitonic complexes. Power-dependent photophysical experiments induce the unconventional decay of X$^-$ together with the superlinear increase of X$^-$ and X$^0$, which hints at a many-body interaction in the confined region at moderate power. Our results open the possibility of sub-diffractional confinement of excitonic complexes for super-resolution imaging, of individual control of localized excitons or trions in semiconducting materials for scalable quantum devices, electrostatically controlled quantum dots coupled to trion spin-photon interfaces, or the realization of efficient platforms to study complicated many-body interaction in nanoscale regions.

## Materials and Methods

### Sample Preparation

Mechanically exfoliated 5-nm thick graphite, 17-nm thick bottom hBN, 10-nm thick top hBN, and a monolayer WSe$_2$ are dry-transferred onto a glass substrate on top of the custom-made PCB. The WSe$_2$ and graphite flakes are in contact with two individual metal electrodes (5-nm Ti followed by 30-nm Au evaporation), which are wire-bonded to the outlet line in the PCB. A conductive AFM probe (Econo-SCM-PIC, Asylum Research) on top of the 3-axis piezoelectric stage (Attocube Inc.) is wire-bonded to the external outlet. Two Keithley 2400 SourceMeters apply the voltage at the back gate ($V_{BG}$) and the AFM tip ($V_{TIP}$) while the sample ($V_S$) is grounded. Both the samples and piezoelectric stages cool down to a nominal temperature of 4 K in the cryostat (Montana Instruments).

**Photoluminescence measurements**

The optical measurement is conducted with a home-built confocal microscopy. A long-working-distance objective lens (Mitutoyo, NA = 0.55) enables optical access to the sample from outside the cryostat through an optically transparent window. Two-axis galvanometer mirrors scan the excitation beam from a HeNe laser source (633 nm) to excite the sample. A tunable continuous-wave laser (M$^2$ SolsTiS) sweeps the wavelength in the PLE experiment. Collected photons are measured by the spectrometer (SP-2500i, Princeton Instruments) or single-photon counting module (Excelitas).

**Acknowledgments**

This work was supported in part by the Army Research Office (ARO) Multidisciplinary University Research Initiative (MURI) program, grant no. W911NF-18-1-0431, in part by National Science Foundation (NSF) Research Advanced by Interdisciplinary Science and Engineering (RAISE), grant no. CHE-1839155. The growth of hexagonal boron nitride crystals was supported by the Elemental Strategy Initiative conducted by the MEXT, Japan, and the CREST (JPMJCR15F3), JST. H.M. acknowledges support by Samsung Scholarship.

## Author Contributions

H.M. and D.E. conceived of the project. *H.M. and L.M. contributed equally to this work. H.M. and L.M. developed and carried out the sample fabrication, optical measurement and analysis, assisted by C.C. and C.P. T.T. and K.W. synthesized hBN crystals. D.E. directed the project. All authors discussed the results and contributed to the manuscript.

## Competing interests

The authors declare no competing interests.

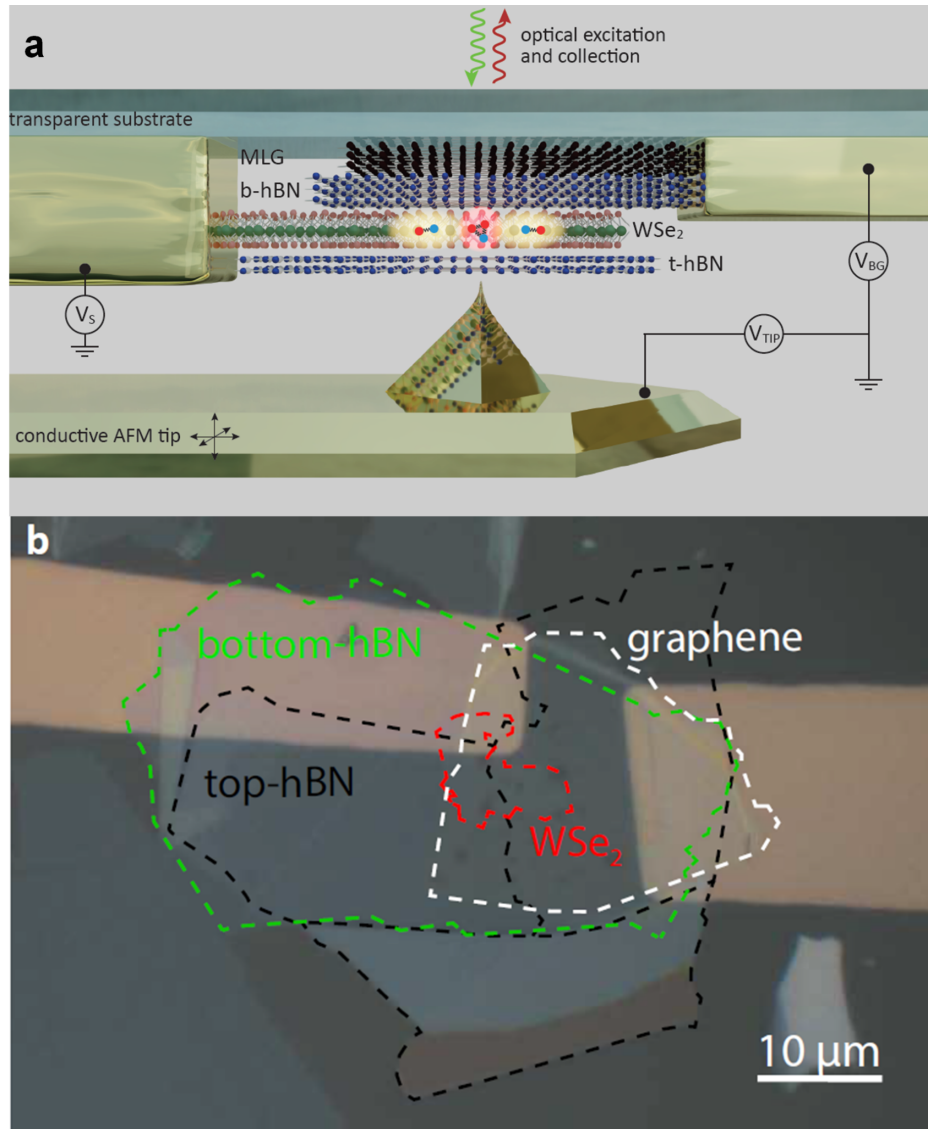

**Fig. 1. Local confinement and control of the excitonic complexes in a monolayer WSe$_2$. a,** A schematic shows the setup for the local density control of the exciton complexes in a two-dimensional semiconductor. A monolayer WSe$_2$, encapsulated by hBN layers and a multilayer graphite back gate, is transferred onto the transparent glass substrate. WSe$_2$ and graphite flakes are connected to individual metal electrodes (V$_S$ and V$_{BG}$). A voltage applied on the conductive AFM tip (V$_{TIP}$), which moves freely on top of the piezoelectric stage, confines the excitonic complexes at the targeted region. **b,** The dashed lines drawn on the optical microscope image show the 2D material flakes and metal electrodes on the transparent glass substrate.

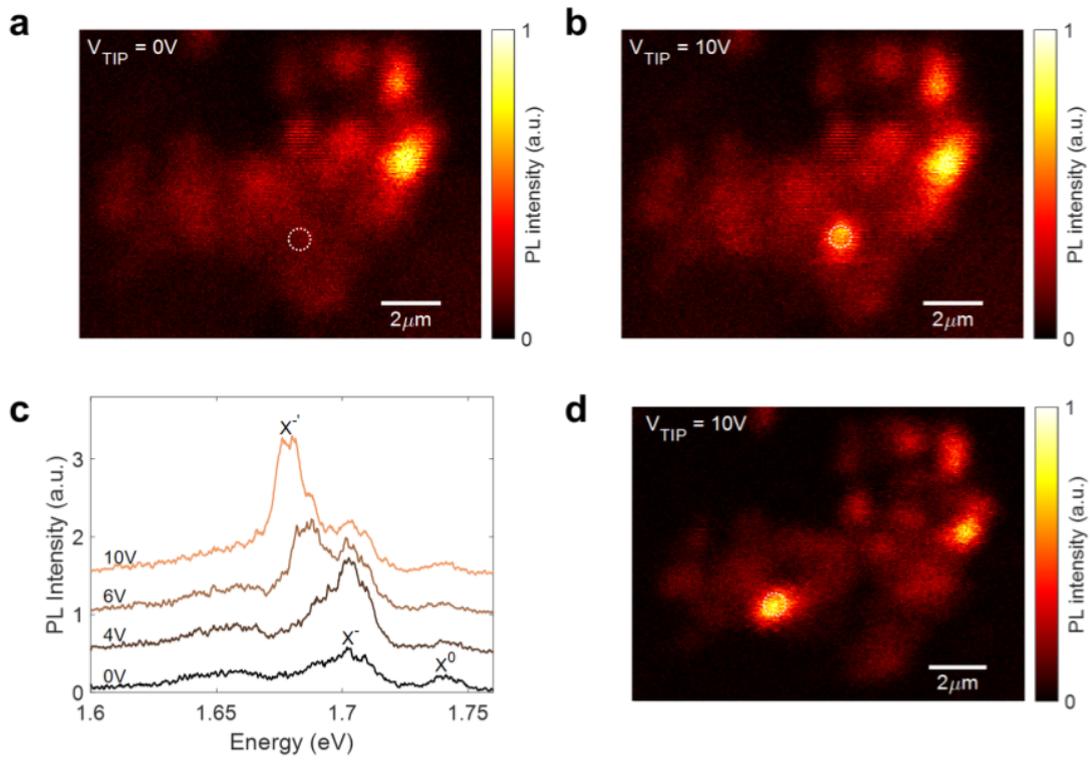

**Fig. 2. Photoluminescence intensity and spectral modulation with the tip voltage. a,** A confocal PL map shows the bright emission of the monolayer WSe$_2$ when $V_{TIP} = V_S = 0$ V and $V_{BG} = -3$ V. **b,** A PL map of the same region when $V_{TIP}$ increases to 10 V. A new bright spot (white circle) appears at the tip position. **c,** Four spectra taken at the center of the bright spot when $V_{TIP}$ increases from 0 V (black) to 10 V (brown) ($V_S = V_{BG} = 0$ V). A new strong peak at 1.675 eV dominates the signal at $V_{TIP} = 10$ V, while the neutral exciton and trion peaks are suppressed. **d,** A PL map of the same region when the tip moves to the other position (white circle). The bright spot moves along the tip position.

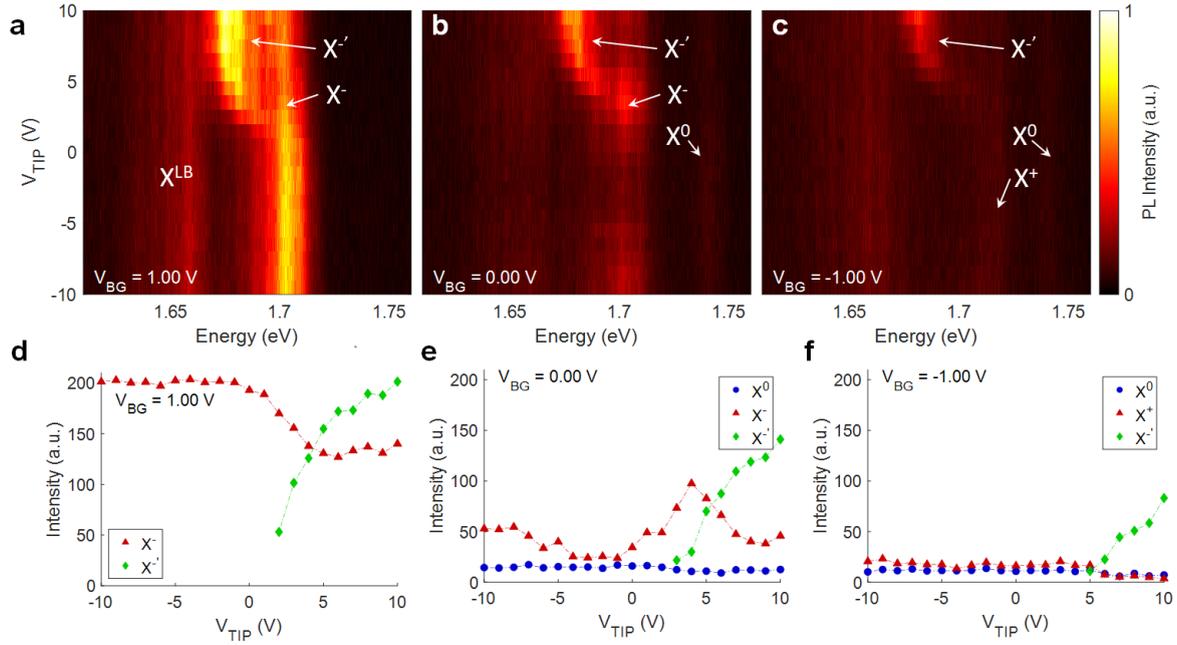

**Fig. 3. Photoluminescence intensity and spectral modulation at different tip and back-gate voltages.** (**a-c**) Spectra at the tip position when $V_{TIP}$ sweeps from -10 to 10 V, at (**a**) $V_{BG}$ = 1 V, (**b**) $V_{BG}$ = 0 V, (**c**) $V_{BG}$ = -1 V. **a,** Trions dominate the signal as the material is negatively doped. $X^{-'}$ peak appears when $V_{TIP}$ > 2 V and becomes stronger as $V_{TIP}$ increases, while the trion peak decreases. **b,** At $V_{BG}$ = 0 V, we observe all of the neutral excitons, negative trions, and $X^{-'}$ peak. The PL emission intensity is weaker than $V_{BG}$ = 1 V as the material is closer to the charge-neutral point. **c,** The onset voltage increases and emission intensity of $X^{-'}$ peak decreases at $V_{BG}$ = -1 V. (**d-f**), The Lorentzian-fitted emission intensity of each peak at (**d**) $V_{BG}$ = 1 V, (**e**) $V_{BG}$ = 0 V, (**f**) $V_{BG}$ = -1 V. As $V_{BG}$ decreases, the material becomes more positively doped, causing the onset voltage of $X^{-'}$ peak to increase and the emission intensity to decrease. However, $X^0$ or $X^-$ emissions are not fully suppressed, even at very high tip voltage, due to the confinement of the electrons.

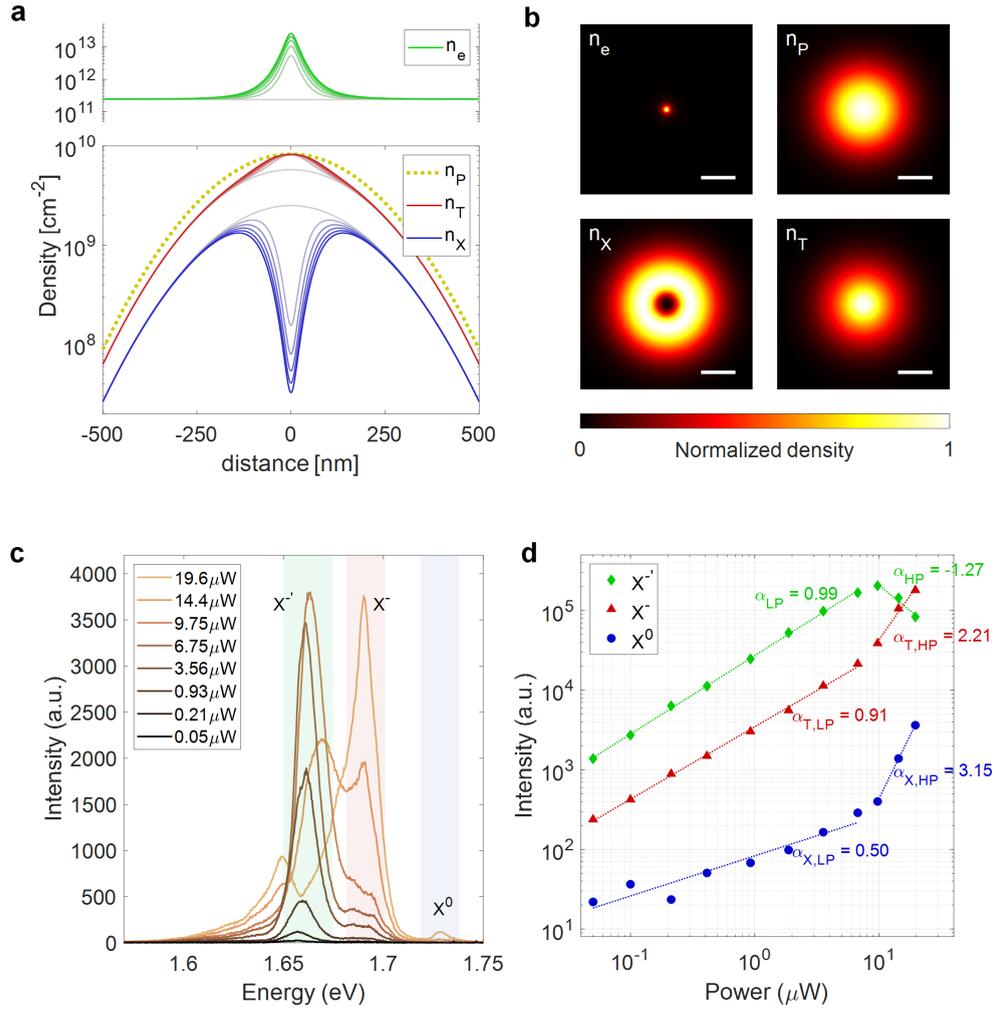

**Fig. 4. Density distribution and power-dependence of excitonic complexes. a,** In-plane distribution of the particle density as $V_{TIP}$ increases from 0 V (gray) to 10 V (vivid) shows subdiffractional confinement of excitonic complexes by local gating ($n_e$: electrons, $n_P$: photo-generated electron-hole pairs, $n_X$: excitons, and $n_T$: trions). **b,** Two-dimensional map of the normalized particle density at $V_{TIP}$ = 10 V, visualizes donut-shaped excitons and confined trions; scale bars are 200 nm. **c,** Power-dependent spectral response of the confined $X^{-'}$ emission which collapses at high power (> 9 $\mu$W) while $X^0$ and $X^-$ peaks appear. **d,** Power-dependent emission intensity of each peak. $X^{-'}$ peak is linearly dependent on the excitation power before the saturation at 9 $\mu$W, then it decreases at higher power. In contrast, excitons and trions are sub-linear at lower power, but they soar up when the $X^{-'}$ peak starts to collapse at high power.

# Supplementary Information for

# Subdiffractional confinement and control of excitonic complexes in a monolayer WSe$_2$

**Estimating density distribution of charged particles**

A conductive AFM tip and WSe$_2$ flake are modeled as a sphere of radius *a*, apart from a conductive plane by distance $d_z$ (Supplementary Fig. 2). The total charge accumulated in the sphere is $Q = \lim_{n\to\infty} \sum_{i=0}^{n} Q_i$, where $Q_{i+1}$ is the effective charge located at position $P_{i+1}$, which keeps the spherical surface equipotential in the presence of the mirror image charge $-Q_i$ at $-P_i$. We stopped the iteration when $Q_i < 10^{-4} Q_0$ to eliminate the smaller terms. An electric potential energy *V(r,z)* and an electric field *E(r,z)* at an arbitrary position can be obtained from this charge distribution, where $r = \sqrt{x^2 + y^2}$ is the radial distance in the WSe$_2$ plane. The capacitance of the system is C = Q/V, and the surface electron density in the WSe$_2$ plane is $\varrho(r) = \varepsilon_0 E(r, z = 0)$, based on Gauss's law.

# Supplementary Figures

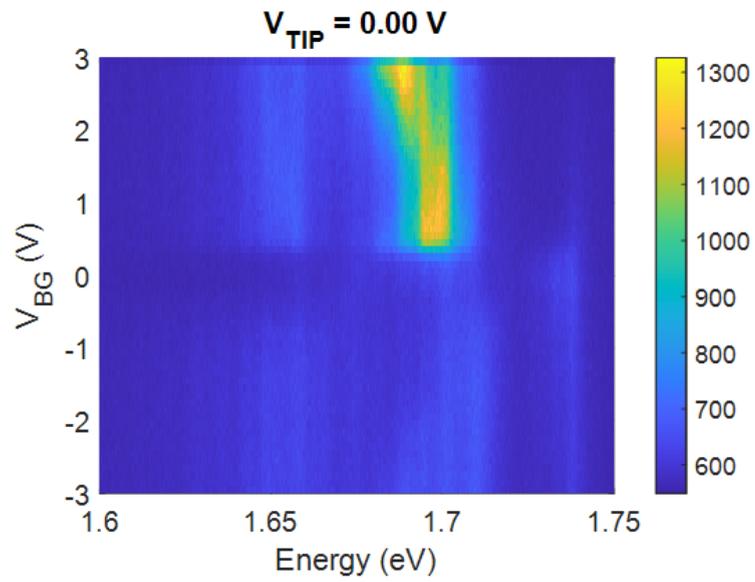

**Supplementary Fig. 1.** Photoluminescence spectra at the same position with Fig. 2a, when $V_{BG}$ sweeps from -3 to 3 V ($V_{TIP}$ = 0 V).

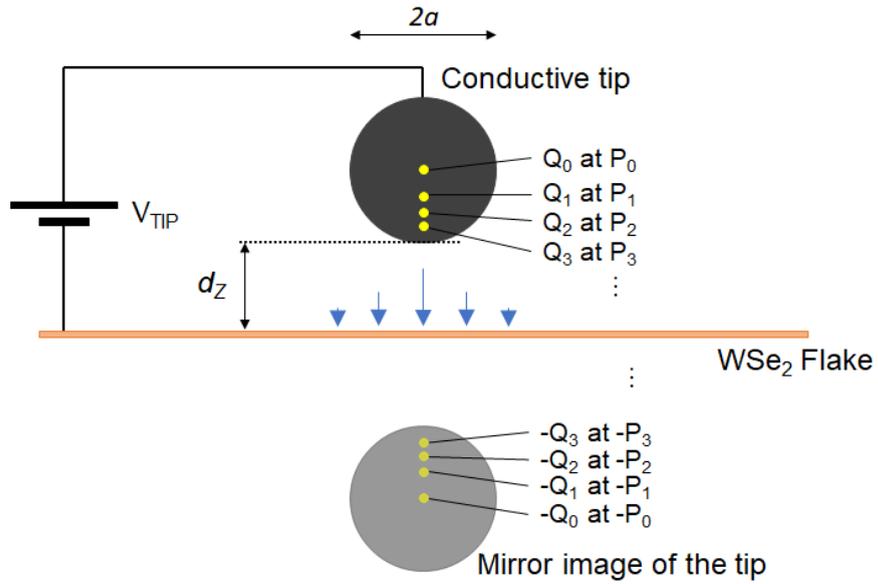

**Supplementary Fig. 2.** Schematics for the charge-distribution calculation. A conductive tip approximated to the sphere of radius $a$, which is away from the WSe$_2$ flake by $d_Z$. The WSe$_2$ flake acts as a conductive plate, and then the local electric field distribution (blue arrows) between the flake and the tip attracts charged carriers in the flake into the central region.

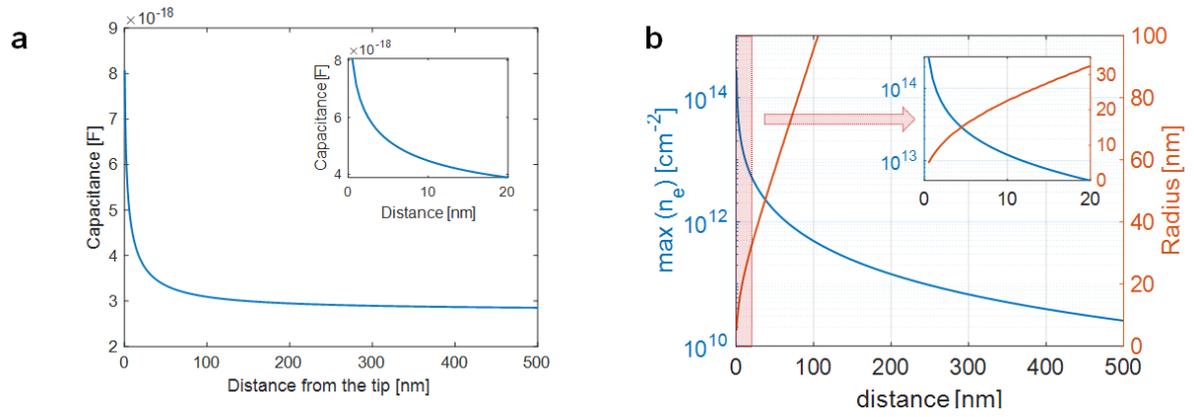

**Supplementary Fig. 3. a,** Calculated capacitance of the proposed system as a function of the vertical distance between the tip and the flake ($d_Z$). **b,** Calculation of maximum density ($n_e$) and confinement radius ($r_C$) of electrons in WSe$_2$ flake as a function of $d_Z$.

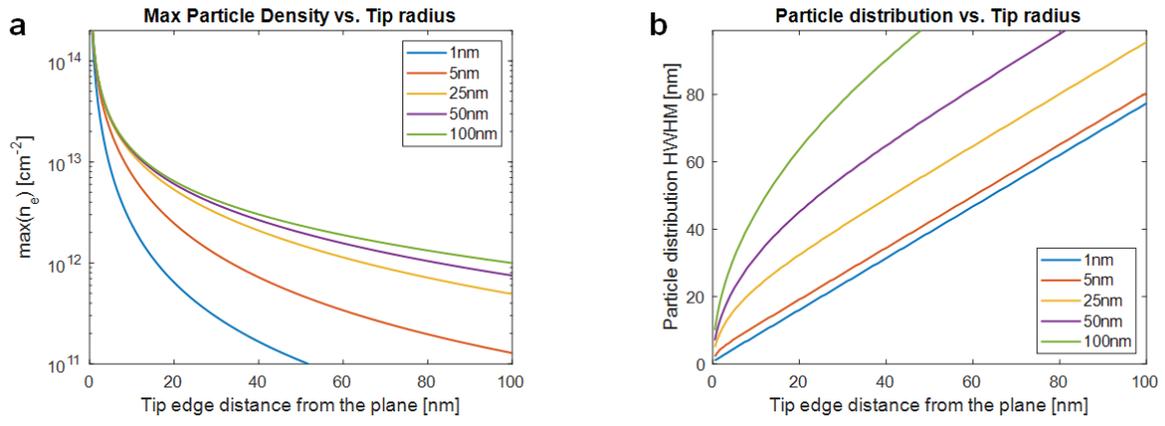

**Supplementary Fig. 4.** Calculation of **a,** maximum charge carrier density ($n_e$) and **b,** confinement radius as a function of $d_Z$, with different tip radiuses ($r_{TIP}$).

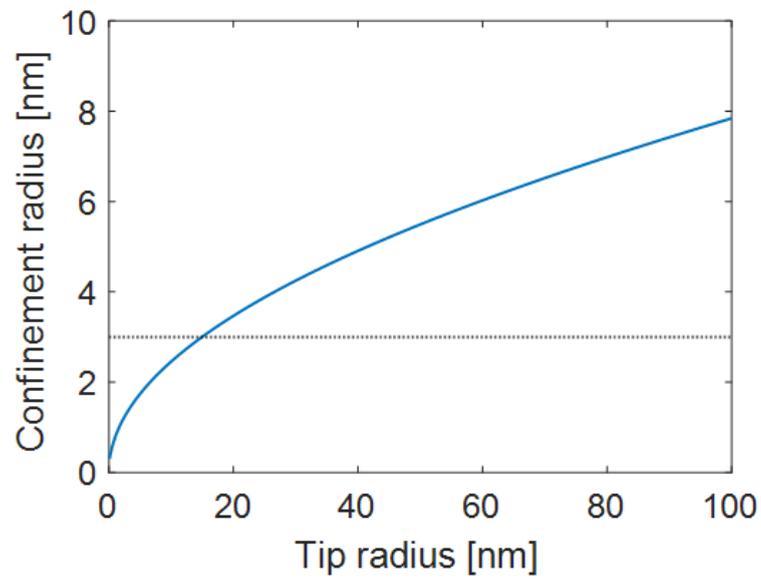

**Supplementary Fig. 5.** Calculation of the confinement radius of electron distribution with a minimum $d_z$ = 0.33 nm (a monolayer hBN). The dotted horizontal line indicates the estimated trion Bohr radius in TMDs.

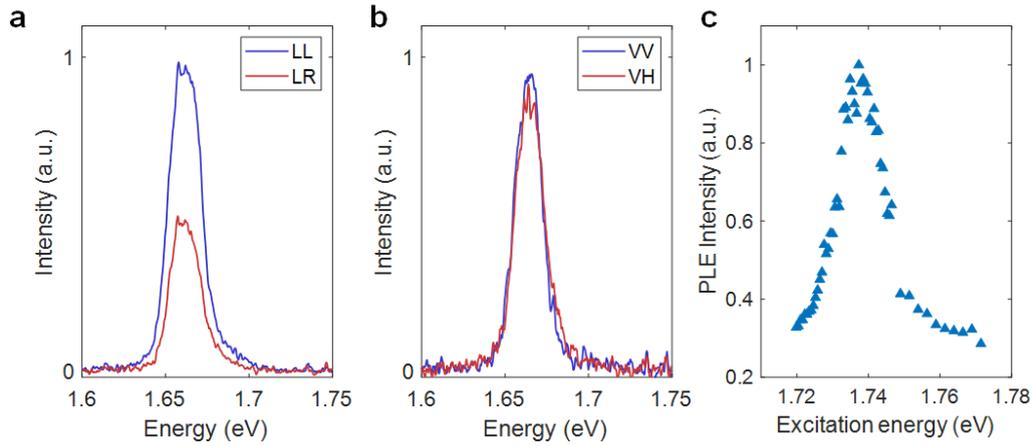

**Supplementary Fig. 6. a,** PL spectra of co-polarization (blue) and cross-polarization (red) when the excitation polarization is circularly polarized. Degree of polarization is 0.37, which shows the valley polarization remains strong for the LP emission process. **b,** PL spectra of co-polarization (blue) and cross-polarization (red) when the excitation polarization is linearly polarized. DOP near to zero are similar to the trion emission in contrast to the exciton emission. **c,** Photoluminescence emission (PLE) spectra as the excitation energy sweeps from 1.77 to 1.72 eV. Strong resonance occurs at the free exciton peak at 1.74 eV.

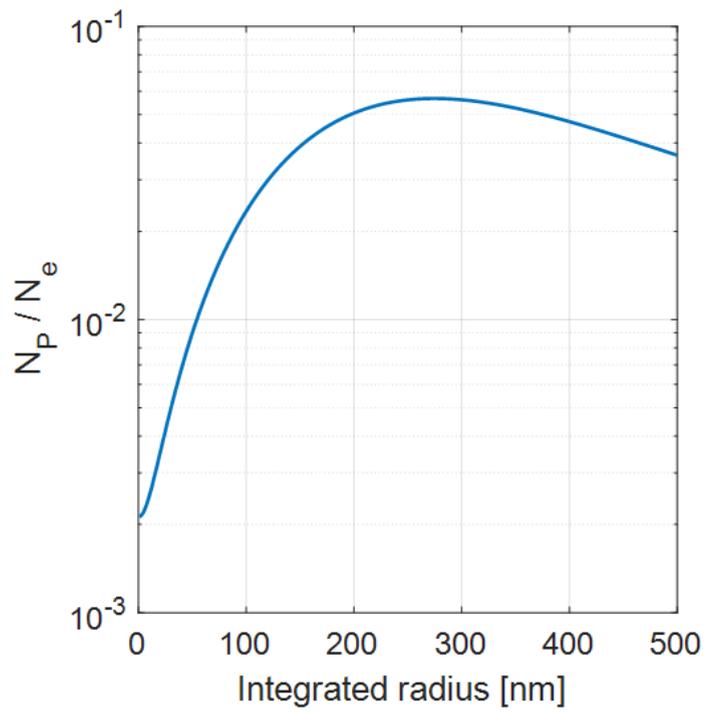

**Supplementary Fig. 7.** The Ratio of the total number of holes ($N_p$) to that of electrons ($N_e$) as a function of integrated radius from the center of the tip.